\documentclass{article}

\usepackage{spconf,amsmath,graphicx,multirow,xcolor,url}
\usepackage[T1]{fontenc}
\usepackage{newtxtext,newtxmath}  % Times text and math, as the ISBI format asks
\usepackage{tikz}
\usetikzlibrary{positioning,arrows.meta}
\usepackage{enumitem}
\usepackage{url}
\usepackage[hidelinks]{hyperref}
\usepackage[hidelinks]{hyperref}

\setlist{nosep, leftmargin=12pt}

\newcommand{\gtad}{G$_{cv}$-TAD}
\newcommand{\res}[4]{\ifcsname res@#1@#2@#3@#4\endcsname\csname res@#1@#2@#3@#4\endcsname\else\textbf{??}\fi}
\newcommand{\gain}[4]{\ifcsname dlt@#1@#2@#3@#4\endcsname\csname dlt@#1@#2@#3@#4\endcsname\else\textbf{??}\fi}
\expandafter\def\csname res@isleeps@teacher@low@macro_f1\endcsname{0.682}
\expandafter\def\csname res@isleeps@teacher@full@macro_f1\endcsname{0.712}
\expandafter\def\csname res@isleeps@eeg@ce@macro_f1\endcsname{0.585}
\expandafter\def\csname res@isleeps@eeg@kd@macro_f1\endcsname{0.626}
\expandafter\def\csname res@isleeps@eeg@gtad_cv@macro_f1\endcsname{0.634}
\expandafter\def\csname res@isleeps@eeg@ce_full@macro_f1\endcsname{0.626}
\expandafter\def\csname res@isleeps@eeg_eog@ce@macro_f1\endcsname{0.617}
\expandafter\def\csname res@isleeps@eeg_eog@kd@macro_f1\endcsname{0.657}
\expandafter\def\csname res@isleeps@eeg_eog@gtad_cv@macro_f1\endcsname{0.675}
\expandafter\def\csname res@isleeps@eeg_eog@ce_full@macro_f1\endcsname{0.666}
\expandafter\def\csname res@isleeps@eog@ce@macro_f1\endcsname{0.567}
\expandafter\def\csname res@isleeps@eog@kd@macro_f1\endcsname{0.626}
\expandafter\def\csname res@isleeps@eog@gtad_cv@macro_f1\endcsname{0.629}
\expandafter\def\csname res@isleeps@eog@ce_full@macro_f1\endcsname{0.642}
\expandafter\def\csname res@isleeps@teacher@low@kappa\endcsname{0.646}
\expandafter\def\csname res@isleeps@teacher@full@kappa\endcsname{0.681}
\expandafter\def\csname res@isleeps@eeg@ce@kappa\endcsname{0.547}
\expandafter\def\csname res@isleeps@eeg@kd@kappa\endcsname{0.597}
\expandafter\def\csname res@isleeps@eeg@gtad_cv@kappa\endcsname{0.596}
\expandafter\def\csname res@isleeps@eeg@ce_full@kappa\endcsname{0.594}
\expandafter\def\csname res@isleeps@eeg_eog@ce@kappa\endcsname{0.584}
\expandafter\def\csname res@isleeps@eeg_eog@kd@kappa\endcsname{0.624}
\expandafter\def\csname res@isleeps@eeg_eog@gtad_cv@kappa\endcsname{0.642}
\expandafter\def\csname res@isleeps@eeg_eog@ce_full@kappa\endcsname{0.635}
\expandafter\def\csname res@isleeps@eog@ce@kappa\endcsname{0.514}
\expandafter\def\csname res@isleeps@eog@kd@kappa\endcsname{0.583}
\expandafter\def\csname res@isleeps@eog@gtad_cv@kappa\endcsname{0.583}
\expandafter\def\csname res@isleeps@eog@ce_full@kappa\endcsname{0.599}
\expandafter\def\csname res@isleeps@teacher@low@f1_N1\endcsname{0.358}
\expandafter\def\csname res@isleeps@teacher@full@f1_N1\endcsname{0.409}
\expandafter\def\csname res@isleeps@eeg@ce@f1_N1\endcsname{0.180}
\expandafter\def\csname res@isleeps@eeg@kd@f1_N1\endcsname{0.233}
\expandafter\def\csname res@isleeps@eeg@gtad_cv@f1_N1\endcsname{0.273}
\expandafter\def\csname res@isleeps@eeg@ce_full@f1_N1\endcsname{0.246}
\expandafter\def\csname res@isleeps@eeg_eog@ce@f1_N1\endcsname{0.241}
\expandafter\def\csname res@isleeps@eeg_eog@kd@f1_N1\endcsname{0.301}
\expandafter\def\csname res@isleeps@eeg_eog@gtad_cv@f1_N1\endcsname{0.331}
\expandafter\def\csname res@isleeps@eeg_eog@ce_full@f1_N1\endcsname{0.305}
\expandafter\def\csname res@isleeps@eog@ce@f1_N1\endcsname{0.192}
\expandafter\def\csname res@isleeps@eog@kd@f1_N1\endcsname{0.249}
\expandafter\def\csname res@isleeps@eog@gtad_cv@f1_N1\endcsname{0.275}
\expandafter\def\csname res@isleeps@eog@ce_full@f1_N1\endcsname{0.301}
\expandafter\def\csname res@isleeps@teacher@low@accuracy\endcsname{0.746}
\expandafter\def\csname res@isleeps@teacher@full@accuracy\endcsname{0.771}
\expandafter\def\csname res@isleeps@eeg@ce@accuracy\endcsname{0.681}
\expandafter\def\csname res@isleeps@eeg@kd@accuracy\endcsname{0.714}
\expandafter\def\csname res@isleeps@eeg@gtad_cv@accuracy\endcsname{0.712}
\expandafter\def\csname res@isleeps@eeg@ce_full@accuracy\endcsname{0.714}
\expandafter\def\csname res@isleeps@eeg_eog@ce@accuracy\endcsname{0.706}
\expandafter\def\csname res@isleeps@eeg_eog@kd@accuracy\endcsname{0.732}
\expandafter\def\csname res@isleeps@eeg_eog@gtad_cv@accuracy\endcsname{0.745}
\expandafter\def\csname res@isleeps@eeg_eog@ce_full@accuracy\endcsname{0.743}
\expandafter\def\csname res@isleeps@eog@ce@accuracy\endcsname{0.659}
\expandafter\def\csname res@isleeps@eog@kd@accuracy\endcsname{0.708}
\expandafter\def\csname res@isleeps@eog@gtad_cv@accuracy\endcsname{0.705}
\expandafter\def\csname res@isleeps@eog@ce_full@accuracy\endcsname{0.718}
\expandafter\def\csname res@naps@teacher@low@macro_f1\endcsname{0.680}
\expandafter\def\csname res@naps@teacher@full@macro_f1\endcsname{0.715}
\expandafter\def\csname res@naps@eeg@ce@macro_f1\endcsname{0.522}
\expandafter\def\csname res@naps@eeg@kd@macro_f1\endcsname{0.587}
\expandafter\def\csname res@naps@eeg@gtad_cv@macro_f1\endcsname{0.601}
\expandafter\def\csname res@naps@eeg@ce_full@macro_f1\endcsname{0.591}
\expandafter\def\csname res@naps@eeg_eog@ce@macro_f1\endcsname{0.588}
\expandafter\def\csname res@naps@eeg_eog@kd@macro_f1\endcsname{0.621}
\expandafter\def\csname res@naps@eeg_eog@gtad_cv@macro_f1\endcsname{0.636}
\expandafter\def\csname res@naps@eeg_eog@ce_full@macro_f1\endcsname{0.661}
\expandafter\def\csname res@naps@eog@ce@macro_f1\endcsname{0.476}
\expandafter\def\csname res@naps@eog@kd@macro_f1\endcsname{0.556}
\expandafter\def\csname res@naps@eog@gtad_cv@macro_f1\endcsname{0.561}
\expandafter\def\csname res@naps@eog@ce_full@macro_f1\endcsname{0.555}
\expandafter\def\csname res@naps@teacher@low@kappa\endcsname{0.668}
\expandafter\def\csname res@naps@teacher@full@kappa\endcsname{0.705}
\expandafter\def\csname res@naps@eeg@ce@kappa\endcsname{0.538}
\expandafter\def\csname res@naps@eeg@kd@kappa\endcsname{0.604}
\expandafter\def\csname res@naps@eeg@gtad_cv@kappa\endcsname{0.613}
\expandafter\def\csname res@naps@eeg@ce_full@kappa\endcsname{0.615}
\expandafter\def\csname res@naps@eeg_eog@ce@kappa\endcsname{0.598}
\expandafter\def\csname res@naps@eeg_eog@kd@kappa\endcsname{0.629}
\expandafter\def\csname res@naps@eeg_eog@gtad_cv@kappa\endcsname{0.645}
\expandafter\def\csname res@naps@eeg_eog@ce_full@kappa\endcsname{0.675}
\expandafter\def\csname res@naps@eog@ce@kappa\endcsname{0.477}
\expandafter\def\csname res@naps@eog@kd@kappa\endcsname{0.557}
\expandafter\def\csname res@naps@eog@gtad_cv@kappa\endcsname{0.558}
\expandafter\def\csname res@naps@eog@ce_full@kappa\endcsname{0.579}
\expandafter\def\csname res@naps@teacher@low@f1_N1\endcsname{0.421}
\expandafter\def\csname res@naps@teacher@full@f1_N1\endcsname{0.444}
\expandafter\def\csname res@naps@eeg@ce@f1_N1\endcsname{0.135}
\expandafter\def\csname res@naps@eeg@kd@f1_N1\endcsname{0.230}
\expandafter\def\csname res@naps@eeg@gtad_cv@f1_N1\endcsname{0.292}
\expandafter\def\csname res@naps@eeg@ce_full@f1_N1\endcsname{0.236}
\expandafter\def\csname res@naps@eeg_eog@ce@f1_N1\endcsname{0.210}
\expandafter\def\csname res@naps@eeg_eog@kd@f1_N1\endcsname{0.288}
\expandafter\def\csname res@naps@eeg_eog@gtad_cv@f1_N1\endcsname{0.324}
\expandafter\def\csname res@naps@eeg_eog@ce_full@f1_N1\endcsname{0.315}
\expandafter\def\csname res@naps@eog@ce@f1_N1\endcsname{0.051}
\expandafter\def\csname res@naps@eog@kd@f1_N1\endcsname{0.147}
\expandafter\def\csname res@naps@eog@gtad_cv@f1_N1\endcsname{0.194}
\expandafter\def\csname res@naps@eog@ce_full@f1_N1\endcsname{0.135}
\expandafter\def\csname res@naps@teacher@low@accuracy\endcsname{0.767}
\expandafter\def\csname res@naps@teacher@full@accuracy\endcsname{0.793}
\expandafter\def\csname res@naps@eeg@ce@accuracy\endcsname{0.685}
\expandafter\def\csname res@naps@eeg@kd@accuracy\endcsname{0.729}
\expandafter\def\csname res@naps@eeg@gtad_cv@accuracy\endcsname{0.733}
\expandafter\def\csname res@naps@eeg@ce_full@accuracy\endcsname{0.737}
\expandafter\def\csname res@naps@eeg_eog@ce@accuracy\endcsname{0.723}
\expandafter\def\csname res@naps@eeg_eog@kd@accuracy\endcsname{0.746}
\expandafter\def\csname res@naps@eeg_eog@gtad_cv@accuracy\endcsname{0.755}
\expandafter\def\csname res@naps@eeg_eog@ce_full@accuracy\endcsname{0.778}
\expandafter\def\csname res@naps@eog@ce@accuracy\endcsname{0.648}
\expandafter\def\csname res@naps@eog@kd@accuracy\endcsname{0.701}
\expandafter\def\csname res@naps@eog@gtad_cv@accuracy\endcsname{0.700}
\expandafter\def\csname res@naps@eog@ce_full@accuracy\endcsname{0.718}
\expandafter\def\csname res@isruc@teacher@low@macro_f1\endcsname{0.788}
\expandafter\def\csname res@isruc@teacher@full@macro_f1\endcsname{0.812}
\expandafter\def\csname res@isruc@eeg@ce@macro_f1\endcsname{0.705}
\expandafter\def\csname res@isruc@eeg@kd@macro_f1\endcsname{0.733}
\expandafter\def\csname res@isruc@eeg@gtad_cv@macro_f1\endcsname{0.733}
\expandafter\def\csname res@isruc@eeg@ce_full@macro_f1\endcsname{0.747}
\expandafter\def\csname res@isruc@eeg_eog@ce@macro_f1\endcsname{0.733}
\expandafter\def\csname res@isruc@eeg_eog@kd@macro_f1\endcsname{0.768}
\expandafter\def\csname res@isruc@eeg_eog@gtad_cv@macro_f1\endcsname{0.772}
\expandafter\def\csname res@isruc@eeg_eog@ce_full@macro_f1\endcsname{0.779}
\expandafter\def\csname res@isruc@eog@ce@macro_f1\endcsname{0.655}
\expandafter\def\csname res@isruc@eog@kd@macro_f1\endcsname{0.702}
\expandafter\def\csname res@isruc@eog@gtad_cv@macro_f1\endcsname{0.711}
\expandafter\def\csname res@isruc@eog@ce_full@macro_f1\endcsname{0.726}
\expandafter\def\csname res@isruc@teacher@low@kappa\endcsname{0.741}
\expandafter\def\csname res@isruc@teacher@full@kappa\endcsname{0.770}
\expandafter\def\csname res@isruc@eeg@ce@kappa\endcsname{0.657}
\expandafter\def\csname res@isruc@eeg@kd@kappa\endcsname{0.686}
\expandafter\def\csname res@isruc@eeg@gtad_cv@kappa\endcsname{0.686}
\expandafter\def\csname res@isruc@eeg@ce_full@kappa\endcsname{0.705}
\expandafter\def\csname res@isruc@eeg_eog@ce@kappa\endcsname{0.689}
\expandafter\def\csname res@isruc@eeg_eog@kd@kappa\endcsname{0.719}
\expandafter\def\csname res@isruc@eeg_eog@gtad_cv@kappa\endcsname{0.728}
\expandafter\def\csname res@isruc@eeg_eog@ce_full@kappa\endcsname{0.741}
\expandafter\def\csname res@isruc@eog@ce@kappa\endcsname{0.598}
\expandafter\def\csname res@isruc@eog@kd@kappa\endcsname{0.645}
\expandafter\def\csname res@isruc@eog@gtad_cv@kappa\endcsname{0.654}
\expandafter\def\csname res@isruc@eog@ce_full@kappa\endcsname{0.675}
\expandafter\def\csname res@isruc@teacher@low@f1_N1\endcsname{0.544}
\expandafter\def\csname res@isruc@teacher@full@f1_N1\endcsname{0.590}
\expandafter\def\csname res@isruc@eeg@ce@f1_N1\endcsname{0.406}
\expandafter\def\csname res@isruc@eeg@kd@f1_N1\endcsname{0.446}
\expandafter\def\csname res@isruc@eeg@gtad_cv@f1_N1\endcsname{0.460}
\expandafter\def\csname res@isruc@eeg@ce_full@f1_N1\endcsname{0.469}
\expandafter\def\csname res@isruc@eeg_eog@ce@f1_N1\endcsname{0.430}
\expandafter\def\csname res@isruc@eeg_eog@kd@f1_N1\endcsname{0.506}
\expandafter\def\csname res@isruc@eeg_eog@gtad_cv@f1_N1\endcsname{0.514}
\expandafter\def\csname res@isruc@eeg_eog@ce_full@f1_N1\endcsname{0.511}
\expandafter\def\csname res@isruc@eog@ce@f1_N1\endcsname{0.321}
\expandafter\def\csname res@isruc@eog@kd@f1_N1\endcsname{0.395}
\expandafter\def\csname res@isruc@eog@gtad_cv@f1_N1\endcsname{0.418}
\expandafter\def\csname res@isruc@eog@ce_full@f1_N1\endcsname{0.432}
\expandafter\def\csname res@isruc@teacher@low@accuracy\endcsname{0.800}
\expandafter\def\csname res@isruc@teacher@full@accuracy\endcsname{0.822}
\expandafter\def\csname res@isruc@eeg@ce@accuracy\endcsname{0.735}
\expandafter\def\csname res@isruc@eeg@kd@accuracy\endcsname{0.759}
\expandafter\def\csname res@isruc@eeg@gtad_cv@accuracy\endcsname{0.758}
\expandafter\def\csname res@isruc@eeg@ce_full@accuracy\endcsname{0.773}
\expandafter\def\csname res@isruc@eeg_eog@ce@accuracy\endcsname{0.760}
\expandafter\def\csname res@isruc@eeg_eog@kd@accuracy\endcsname{0.783}
\expandafter\def\csname res@isruc@eeg_eog@gtad_cv@accuracy\endcsname{0.790}
\expandafter\def\csname res@isruc@eeg_eog@ce_full@accuracy\endcsname{0.801}
\expandafter\def\csname res@isruc@eog@ce@accuracy\endcsname{0.694}
\expandafter\def\csname res@isruc@eog@kd@accuracy\endcsname{0.728}
\expandafter\def\csname res@isruc@eog@gtad_cv@accuracy\endcsname{0.734}
\expandafter\def\csname res@isruc@eog@ce_full@accuracy\endcsname{0.751}
\expandafter\def\csname dlt@isleeps@eeg@gtad_cv@ce\endcsname{+5.4}
\expandafter\def\csname dlt@isleeps@eeg@gtad_cv@kd\endcsname{+0.5}
\expandafter\def\csname dlt@isleeps@eeg@kd@ce\endcsname{+4.9}
\expandafter\def\csname dlt@isleeps@eeg@gtad_cv@ce_full\endcsname{+0.8}
\expandafter\def\csname dlt@isleeps@eeg@kd@ce_full\endcsname{+0.3}
\expandafter\def\csname dlt@isleeps@eeg_eog@gtad_cv@ce\endcsname{+5.9}
\expandafter\def\csname dlt@isleeps@eeg_eog@gtad_cv@kd\endcsname{+1.6}
\expandafter\def\csname dlt@isleeps@eeg_eog@kd@ce\endcsname{+4.4}
\expandafter\def\csname dlt@isleeps@eeg_eog@gtad_cv@ce_full\endcsname{+0.9}
\expandafter\def\csname dlt@isleeps@eeg_eog@kd@ce_full\endcsname{$-$0.7}
\expandafter\def\csname dlt@isleeps@eog@gtad_cv@ce\endcsname{+6.9}
\expandafter\def\csname dlt@isleeps@eog@gtad_cv@kd\endcsname{+0.4}
\expandafter\def\csname dlt@isleeps@eog@kd@ce\endcsname{+6.5}
\expandafter\def\csname dlt@isleeps@eog@gtad_cv@ce_full\endcsname{$-$1.4}
\expandafter\def\csname dlt@isleeps@eog@kd@ce_full\endcsname{$-$1.7}
\expandafter\def\csname dlt@naps@eeg@gtad_cv@ce\endcsname{+6.8}
\expandafter\def\csname dlt@naps@eeg@gtad_cv@kd\endcsname{+1.2}
\expandafter\def\csname dlt@naps@eeg@kd@ce\endcsname{+5.6}
\expandafter\def\csname dlt@naps@eeg@gtad_cv@ce_full\endcsname{+0.5}
\expandafter\def\csname dlt@naps@eeg@kd@ce_full\endcsname{$-$0.8}
\expandafter\def\csname dlt@naps@eeg_eog@gtad_cv@ce\endcsname{+4.9}
\expandafter\def\csname dlt@naps@eeg_eog@gtad_cv@kd\endcsname{+1.5}
\expandafter\def\csname dlt@naps@eeg_eog@kd@ce\endcsname{+3.4}
\expandafter\def\csname dlt@naps@eeg_eog@gtad_cv@ce_full\endcsname{$-$1.5}
\expandafter\def\csname dlt@naps@eeg_eog@kd@ce_full\endcsname{$-$3.0}
\expandafter\def\csname dlt@naps@eog@gtad_cv@ce\endcsname{+7.6}
\expandafter\def\csname dlt@naps@eog@gtad_cv@kd\endcsname{+0.9}
\expandafter\def\csname dlt@naps@eog@kd@ce\endcsname{+6.7}
\expandafter\def\csname dlt@naps@eog@gtad_cv@ce_full\endcsname{+0.3}
\expandafter\def\csname dlt@naps@eog@kd@ce_full\endcsname{$-$0.6}
\expandafter\def\csname dlt@isruc@eeg@gtad_cv@ce\endcsname{+3.2}
\expandafter\def\csname dlt@isruc@eeg@gtad_cv@kd\endcsname{+0.3}
\expandafter\def\csname dlt@isruc@eeg@kd@ce\endcsname{+3.0}
\expandafter\def\csname dlt@isruc@eeg@gtad_cv@ce_full\endcsname{$-$1.7}
\expandafter\def\csname dlt@isruc@eeg@kd@ce_full\endcsname{$-$2.0}
\expandafter\def\csname dlt@isruc@eeg_eog@gtad_cv@ce\endcsname{+4.2}
\expandafter\def\csname dlt@isruc@eeg_eog@gtad_cv@kd\endcsname{+0.6}
\expandafter\def\csname dlt@isruc@eeg_eog@kd@ce\endcsname{+3.6}
\expandafter\def\csname dlt@isruc@eeg_eog@gtad_cv@ce_full\endcsname{$-$0.9}
\expandafter\def\csname dlt@isruc@eeg_eog@kd@ce_full\endcsname{$-$1.6}
\expandafter\def\csname dlt@isruc@eog@gtad_cv@ce\endcsname{+5.7}
\expandafter\def\csname dlt@isruc@eog@gtad_cv@kd\endcsname{+1.0}
\expandafter\def\csname dlt@isruc@eog@kd@ce\endcsname{+4.7}
\expandafter\def\csname dlt@isruc@eog@gtad_cv@ce_full\endcsname{$-$2.0}
\expandafter\def\csname dlt@isruc@eog@kd@ce_full\endcsname{$-$3.0}
\newcommand{\frag}[2]{\ifcsname frag@#1@#2\endcsname\csname frag@#1@#2\endcsname\else\textbf{??}\fi}
\expandafter\def\csname frag@sleepedf20@awakenings_per_h\endcsname{1.6}
\expandafter\def\csname frag@sleepedf20@shifts_per_h\endcsname{13.9}
\expandafter\def\csname frag@sleepedf20@n1_pct\endcsname{8.2}
\expandafter\def\csname frag@sleepedf20@sleep_eff\endcsname{91.8}
\expandafter\def\csname frag@isleeps@awakenings_per_h\endcsname{3.1}
\expandafter\def\csname frag@isleeps@shifts_per_h\endcsname{12.4}
\expandafter\def\csname frag@isleeps@n1_pct\endcsname{14.3}
\expandafter\def\csname frag@isleeps@sleep_eff\endcsname{79.6}
\expandafter\def\csname frag@naps@awakenings_per_h\endcsname{3.8}
\expandafter\def\csname frag@naps@shifts_per_h\endcsname{17.6}
\expandafter\def\csname frag@naps@n1_pct\endcsname{13.8}
\expandafter\def\csname frag@naps@sleep_eff\endcsname{81.1}
\expandafter\def\csname frag@isruc@awakenings_per_h\endcsname{4.5}
\expandafter\def\csname frag@isruc@shifts_per_h\endcsname{22.7}
\expandafter\def\csname frag@isruc@n1_pct\endcsname{17.2}
\expandafter\def\csname frag@isruc@sleep_eff\endcsname{83.0}
\expandafter\def\csname eff@teacher@params\endcsname{136{,}155{,}077}
\expandafter\def\csname eff@teacher@fp32_kb\endcsname{544620.3}
\expandafter\def\csname eff@teacher@int8_kb\endcsname{204332}
\expandafter\def\csname eff@teacher@mmacs\endcsname{10381.3}
\expandafter\def\csname eff@teacher@ms\endcsname{271.90}
\expandafter\def\csname eff@teacher@gmacs\endcsname{10.4}
\expandafter\def\csname eff@teacher@ratio\endcsname{1}
\expandafter\def\csname eff@student_1ch@params\endcsname{30{,}421}
\expandafter\def\csname eff@student_1ch@fp32_kb\endcsname{121.7}
\expandafter\def\csname eff@student_1ch@int8_kb\endcsname{75}
\expandafter\def\csname eff@student_1ch@mmacs\endcsname{2.2}
\expandafter\def\csname eff@student_1ch@ms\endcsname{1.16}
\expandafter\def\csname eff@student_1ch@gmacs\endcsname{0.0}
\expandafter\def\csname eff@student_1ch@ratio\endcsname{4{,}476}
\expandafter\def\csname eff@student_2ch@params\endcsname{31{,}221}
\expandafter\def\csname eff@student_2ch@fp32_kb\endcsname{124.9}
\expandafter\def\csname eff@student_2ch@int8_kb\endcsname{78}
\expandafter\def\csname eff@student_2ch@mmacs\endcsname{2.7}
\expandafter\def\csname eff@student_2ch@ms\endcsname{1.22}
\expandafter\def\csname eff@student_2ch@gmacs\endcsname{0.0}
\expandafter\def\csname eff@student_2ch@ratio\endcsname{4{,}361}
\newcommand{\eff}[2]{\ifcsname eff@#1@#2\endcsname\csname eff@#1@#2\endcsname\else\textbf{??}\fi}

\newcommand{\nrecall}[2]{\csname nrec@#1@#2\endcsname}
\expandafter\def\csname nrec@isleeps@ce\endcsname{19}
\expandafter\def\csname nrec@isleeps@gtad_cv\endcsname{31}
\expandafter\def\csname nrec@isleeps@ce_full\endcsname{25}
\expandafter\def\csname nrec@naps@ce\endcsname{15}
\expandafter\def\csname nrec@naps@gtad_cv\endcsname{29}
\expandafter\def\csname nrec@naps@ce_full\endcsname{24}
\expandafter\def\csname nrec@isruc@ce\endcsname{39}
\expandafter\def\csname nrec@isruc@gtad_cv\endcsname{52}
\expandafter\def\csname nrec@isruc@ce_full\endcsname{47}

\title{Towards wearable-scale, transition-aware sleep staging\\
for clinical populations}

\name{Suvadeep Maiti$^{\ddagger}$,
Keya Sinha$^{\star,\dagger}$,
Raju S. Bapi$^{\star}$}

\address{
$^{\ddagger}$University College London, London, United Kingdom\\
$^{\dagger}$Indian Institute of Science Education and Research
(IISER) Bhopal, India\\
$^{\star}$IHub-Data, International Institute of Information Technology,
Hyderabad, India
}

\begin{document}
\maketitle

\begin{abstract}

Automatic sleep staging models are predominantly developed on healthy sleepers and generalize poorly to clinical populations with fragmented sleep. Across three cohorts (patients with stroke, isolated REM sleep behaviour disorder, and other sleep disorders), we observe 2--3$\times$ more awakenings per hour, nearly twice as much N1, and less REM sleep than in healthy sleepers. We propose a label-efficient teacher--student framework for wearable-scale sleep staging. SleepGPT (136M parameters), adapted using labels from 25\% of training patients, supervises a causal CNN-GRU (31k parameters; 122~KB) that stages EEG, EOG, or both in real time. We introduce \gtad, a new black-box transition-aware distillation loss. It emphasizes uncommon stage transitions while discounting those that vary across patients, using a prior estimated from labelled patients. In patient-wise 10-fold cross-validation, \gtad\ improves per-patient macro F1 over supervised training on the same labels by
\gain{isruc}{eeg}{gtad_cv}{ce}--\gain{naps}{eog}{gtad_cv}{ce} points on average. Gains are largest on N1, and performance approaches full-label training. 
A summary of the paper and the code are available at
\url{https://github.com/suvadeepmaiti/clinical-sleep-staging-gcv-tad}.

\end{abstract}

% \begin{keywords}
% Sleep staging, wearable, knowledge distillation, foundation models, label efficiency, stroke, REM sleep behaviour disorder
% \end{keywords}

\section{Introduction}
\label{sec:intro}

Automatic sleep staging is essential for assessing sleep disorders, but manual polysomnography (PSG) scoring is labor-intensive and unsuitable for long-term monitoring. Wearable devices using one or two EEG/EOG channels could enable continuous sleep assessment outside the laboratory. However, most staging models are developed on healthy or community cohorts \cite{supratak2017deepsleepnet,perslev2021usleep} and often generalize poorly to clinical populations. Recent work demonstrated substantial performance degradation when a model trained on healthy sleepers was evaluated on ischemic stroke patients \cite{bose2026isleeps}. Clinical sleep is also more fragmented, with frequent awakenings, increased N1 sleep, and heterogeneous stage-transition patterns. These differences make staging more challenging, while the scarcity of expert-labelled recordings limits patient-specific model adaptation.

Recent sleep foundation models, such as SleepGPT \cite{huang2026sleepgpt}, offer powerful representations but require substantial memory and bidirectional context, limiting their use in real-time wearable applications. Knowledge distillation (KD) \cite{hinton2015distilling} can transfer knowledge from large models to smaller students. However, conventional KD treats epochs independently and does not explicitly account for sleep-stage transitions. Existing sleep-specific approaches primarily transfer intermediate representations using teacher assistants \cite{liang2023teacherassistant,jia2024distillsleepnet}. We instead propose a black-box distillation framework that incorporates the transition dynamics of the target clinical population while requiring labels from only a fraction of patients.

 \textbf{Our contributions are:} \textbf{(1)} A novel transition-aware distillation loss, \gtad, that emphasizes uncommon stage transitions while accounting for between-patient variability. \textbf{(2)} A lightweight causal CNN-GRU (31k parameters; 122~KB) distilled from SleepGPT using only 25\% labelled training patients, enabling real-time sleep staging from EEG and/or EOG. \textbf{(3)} A comprehensive evaluation across three clinical cohorts using patient-wise 10-fold cross-validation, demonstrating improved macro F1 and N1 staging over supervised training with the same labels and performance approaching full-label training.

\section{Method}
\label{sec:method}

\begin{figure*}[t]
    \centering
    \includegraphics[width=\textwidth]{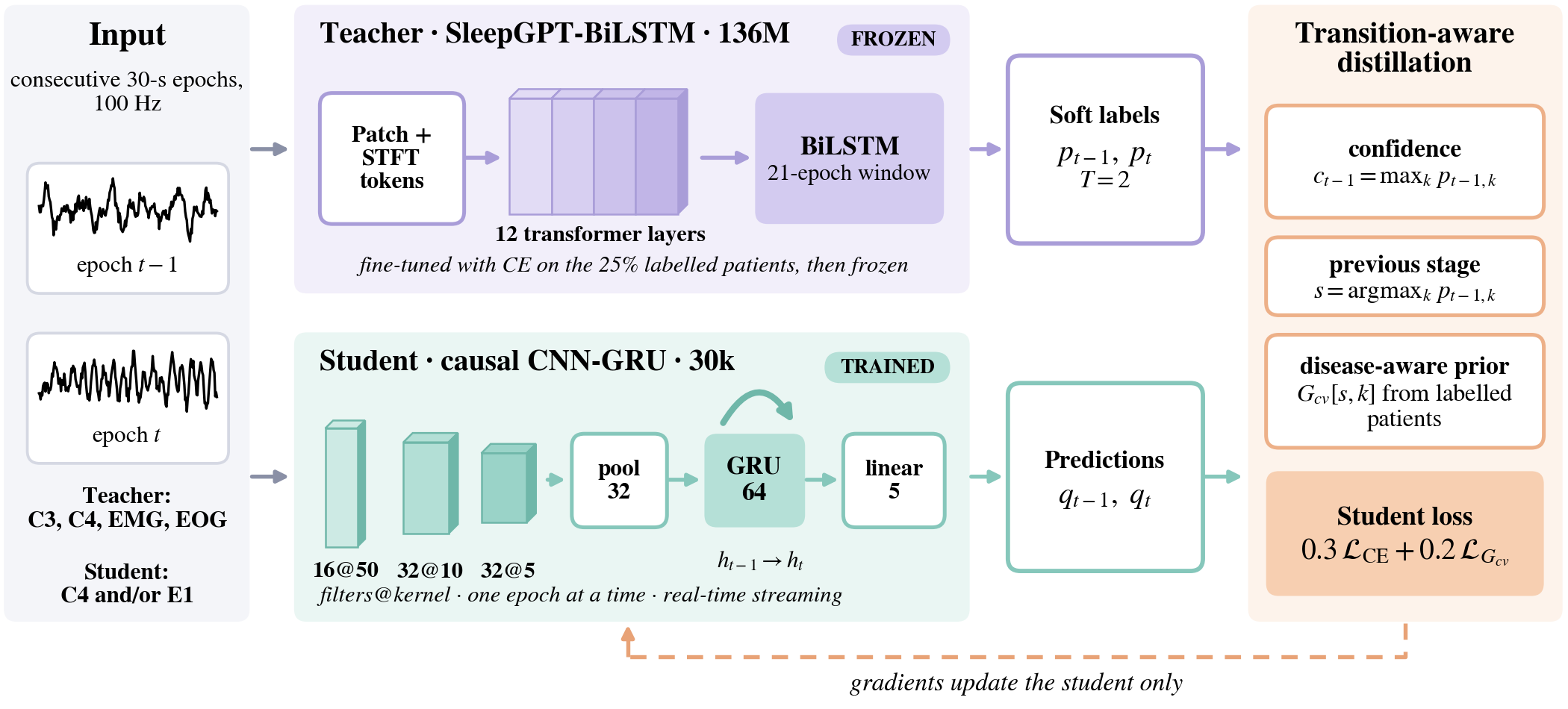}
    \caption{Overview of the proposed \gtad\ framework.
    The teacher is adapted using 25\% labelled patients and
    frozen during distillation. Its predictions guide a causal
    student through a transition-aware loss weighted by
    preceding-stage confidence and a cohort-specific prior.
    Only the student is intended for deployment.}
    \label{fig:pipeline}
\end{figure*}

\textbf{Teacher:}
We use SleepGPT \cite{huang2026sleepgpt}, a 136M-parameter transformer pretrained on 86k~h of PSG, followed by a 3-layer BiLSTM (100 hidden units). The model processes four PSG channels (C3, C4, EMG, EOG) using a 21-epoch context to classify five sleep stages. For each clinical cohort, we independently fine-tune the teacher using 25\% and 100\% of the labelled training patients. The fully labelled teacher serves as a reference, while the 25\%-label teacher provides supervision for student distillation. Fine-tuning uses cross-entropy (CE) and AdamW ($10^{-4}$ learning rate, 0.75 layer decay, at most 25 epochs), with model selection based on inner-validation macro F1. The frozen 25\%-label teacher generates soft predictions $p_t=\mathrm{softmax}(z_t^T/T)$ with $T=2$ and
hard pseudo-labels $\hat y_t=\arg\max_k p_{t,k}$
for all training epochs.

\textbf{Student:}
We design a lightweight causal CNN-GRU with three 1-D convolutional layers (16/32/32 filters; kernel sizes
50/10/5), followed by adaptive pooling to 32 steps, a 64-unit GRU, and a five-class classifier. The model contains 30,421 parameters for a single channel and 31,221 for EEG+EOG. By preserving the GRU hidden state across consecutive 30-s epochs, the student performs streaming sleep staging without future context. It is trained on 20-epoch sequences using Adam (learning rate $10^{-3}$, cosine schedule, maximum 40 epochs), with the best checkpoint selected by inner-validation macro F1.

\textbf{Transition-aware distillation:}
Clinical sleep is characterized by frequent and heterogeneous
stage transitions, making it important to account for
transition dynamics during knowledge transfer.
To address this, we propose \gtad, a black-box
distillation loss that incorporates cohort-specific
transition statistics and between-patient variability
to guide student learning.

Let $p_t$ and $q_t$ denote the teacher and student
probabilities at epoch $t$, respectively, with temperature
$T=2$. The teacher's preceding prediction defines
$s_{t-1}=\arg\max_k p_{t-1,k}$ and confidence
$c_{t-1}=\max_k p_{t-1,k}$.
Our transition-aware loss is

\begin{equation}
\begin{aligned}
\mathcal{L}_{G_{cv}}
&=\frac{T^2}{L-1}\sum_{t=2}^{L}c_{t-1}\\
&\quad\cdot\sum_{k=1}^{5}G_{cv}[s_{t-1},k]\,
d(p_{t,k},q_{t,k}),
\end{aligned}
\label{eq:lg}
\end{equation}
where $d(p,q)=p\log(p/q)-p+q$ is a non-negative
generalized KL term. The transition prior $G_{cv}$
is defined as

\begin{equation}
G_{cv}=\frac{1-\bar M}{1+\lambda\,\mathrm{CV}},
\qquad \lambda=0.5,
\label{eq:gcv}
\end{equation}
where $\bar M$ is the mean of the per-patient
row-normalized transition matrices and
$\mathrm{CV}=\sigma/(\bar M+\epsilon)$ measures their
between-patient variability. Here, $\sigma$ denotes
the element-wise standard deviation.

The prior assigns greater weight to less probable
next-stage transitions while discounting those that
vary strongly between patients. It is estimated
separately for each fold using only labelled training
patients. The student minimizes
$0.3\mathcal{L}_{CE}+0.2\mathcal{L}_{G_{cv}}$,
where CE uses scorer labels for labelled patients
and teacher pseudo-labels for unlabelled patients.

\section{Experiments}
\label{sec:exp}
% ---------------------------------------------------------------------------
% Table: cohorts (inlined from tables/cohorts.tex)
% ---------------------------------------------------------------------------

\textbf{Datasets:}
We evaluate three clinical polysomnography (PSG) datasets. \emph{iSLEEPS}
\cite{maiti2026isleepsdata} contains overnight PSG
recordings from 100 ischemic stroke patients.
\emph{NAPS} \cite{zhang2018nsrr, elliott2023baseline, lim2026north}
includes patients with isolated REM sleep behaviour disorder (iRBD), recorded across multiple clinical sites and centrally scored using AASM guidelines. We retain 133 of the 138 recordings that contain the required central EEG, chin EMG, and EOG channels. \emph{ISRUC-S1} \cite{khalighi2016isruc} comprises patients with various sleep disorders. We use 99 recordings with scorer-1 annotations, excluding recording 40 due to its different montage.

\begin{table*}[t]
\centering
\caption{Test accuracy (Acc), macro F1 (MF1) and Cohen's $\kappa$ (mean over 10 folds) of models trained on 100\% of the labels (CE) and with 25\% of the training patients labelled (CE or G$_{cv}$-TAD), and per-class F1 with 25\% labels (teacher: CE; students: G$_{cv}$-TAD). Teacher: 4-channel SleepGPT-BiLSTM; students: 31k-parameter causal CNN-GRU. Best G$_{cv}$-TAD student of each cohort in bold.}
\label{tab:main}
\footnotesize
\setlength{\tabcolsep}{3pt}
\begin{tabular}{ll|ccc|ccc|ccc|ccccc}
\hline
 & & \multicolumn{3}{c|}{100\% labels} & \multicolumn{6}{c|}{25\% labelled patients} & \multicolumn{5}{c}{Per-class F1, 25\%} \\
 & & \multicolumn{3}{c|}{CE} & \multicolumn{3}{c|}{CE} & \multicolumn{3}{c|}{G$_{cv}$-TAD} & & & & & \\
Cohort & Model & Acc & MF1 & $\kappa$ & Acc & MF1 & $\kappa$ & Acc & MF1 & $\kappa$ & W & N1 & N2 & N3 & REM \\
\hline
\multirow{4}{*}{iSLEEPS} & Teacher & 0.77 & 0.71 & 0.68 & 0.75 & 0.68 & 0.65 & -- & -- & -- & 0.81 & 0.36 & 0.80 & 0.70 & 0.74 \\
 & EOG (E1) & 0.72 & 0.64 & 0.60 & 0.66 & 0.57 & 0.51 & 0.71 & 0.63 & 0.58 & 0.76 & 0.28 & 0.77 & 0.68 & 0.66 \\
 & EEG (C4) & 0.71 & 0.63 & 0.59 & 0.68 & 0.59 & 0.55 & 0.71 & 0.63 & 0.60 & 0.77 & 0.27 & 0.78 & 0.69 & 0.66 \\
 & EEG+EOG & 0.74 & 0.67 & 0.64 & 0.71 & 0.62 & 0.58 & \textbf{0.75} & \textbf{0.68} & \textbf{0.64} & \textbf{0.80} & \textbf{0.33} & \textbf{0.81} & \textbf{0.71} & \textbf{0.73} \\
\hline
\multirow{4}{*}{NAPS} & Teacher & 0.79 & 0.71 & 0.70 & 0.77 & 0.68 & 0.67 & -- & -- & -- & 0.84 & 0.42 & 0.82 & 0.58 & 0.73 \\
 & EOG (E1) & 0.72 & 0.56 & 0.58 & 0.65 & 0.48 & 0.48 & 0.70 & 0.56 & 0.56 & 0.77 & 0.19 & 0.78 & 0.46 & 0.60 \\
 & EEG (C4) & 0.74 & 0.59 & 0.61 & 0.68 & 0.52 & 0.54 & 0.73 & 0.60 & 0.61 & 0.82 & 0.29 & 0.81 & 0.45 & 0.63 \\
 & EEG+EOG & 0.78 & 0.66 & 0.67 & 0.72 & 0.59 & 0.60 & \textbf{0.76} & \textbf{0.64} & \textbf{0.64} & \textbf{0.84} & \textbf{0.32} & \textbf{0.82} & \textbf{0.51} & \textbf{0.70} \\
\hline
\multirow{4}{*}{ISRUC-S1} & Teacher & 0.82 & 0.81 & 0.77 & 0.80 & 0.79 & 0.74 & -- & -- & -- & 0.88 & 0.54 & 0.79 & 0.87 & 0.87 \\
 & EOG (E1) & 0.75 & 0.73 & 0.68 & 0.69 & 0.66 & 0.60 & 0.73 & 0.71 & 0.65 & 0.80 & 0.42 & 0.73 & 0.84 & 0.76 \\
 & EEG (C4) & 0.77 & 0.75 & 0.71 & 0.73 & 0.70 & 0.66 & 0.76 & 0.73 & 0.69 & 0.83 & 0.46 & 0.77 & 0.85 & 0.76 \\
 & EEG+EOG & 0.80 & 0.78 & 0.74 & 0.76 & 0.73 & 0.69 & \textbf{0.79} & \textbf{0.77} & \textbf{0.73} & \textbf{0.86} & \textbf{0.51} & \textbf{0.79} & \textbf{0.87} & \textbf{0.83} \\
\hline
\end{tabular}
\end{table*}

\textbf{Evaluation protocol:}
We use patient-wise 10-fold cross-validation with
80:10:10 training, validation, and test splits across all three cohorts. Teachers and students are independently trained with CE using 100\% or 25\% of training labels. The 100\%-label models serve as supervised references. For distillation, the frozen 25\%-label teacher supervises students using KD or \gtad, with the remaining training patients treated as unlabelled. All methods share identical splits, and checkpoints are selected using validation macro F1. Test patients remain held out throughout training and selection. We report accuracy, macro F1, Cohen's $\kappa$, and per-stage F1. 

% Significance is assessed using two-sided Wilcoxon signed-rank tests on per patient macro F1, with Holm correction across nine settings.

% ---------------------------------------------------------------------------
% Table: main (inlined from tables/main.tex)
% ---------------------------------------------------------------------------

\section{Results and Discussion}
\label{sec:results}

\begin{table}[t]
\centering
\caption{iSLEEPS and ISRUC-S1 against published results
(100\% labels unless marked 25\%; Ch.: input channels).
Ours: patient-wise 10-fold CV, fold means. Published
protocols (folds, recordings) differ from ours.
Best per cohort in bold.}
\label{tab:isruc}
\footnotesize
\setlength{\tabcolsep}{2.2pt}
\begin{tabular}{lcrccc}
\hline
Method & Ch. & Params & Acc & MF1 & $\kappa$ \\
\hline
\multicolumn{6}{l}{\textit{iSLEEPS}} \\
SE-ResNet-BiLSTM$^{a}$ \cite{bose2026isleeps}
& 1 & -- & 0.75 & 0.68 & 0.64 \\
SleepStageNet \cite{alhazmi2026sleepstagenet}
& 1 & 1.1M & 0.72 & 0.64 & 0.60 \\
SleepStageNet \cite{alhazmi2026sleepstagenet}
& 5 & 1.1M & 0.74 & 0.67 & 0.63 \\
Ours: teacher
& 4 & 136.2M & \textbf{0.77} & \textbf{0.71}
& \textbf{0.68} \\
Ours: student, \gtad, 25\%
& 2 & 31k & 0.75 & 0.68 & 0.64 \\
\hline
\multicolumn{6}{l}{\textit{ISRUC-S1}} \\
MSSC-BiMamba$^{b}$ \cite{zhang2024mssc}
& 10 & 470k & \textbf{0.83} & 0.80
& \textbf{0.77} \\
JK-STGCN \cite{ji2022jkstgcn}
& -- & -- & 0.82 & 0.80 & \textbf{0.77} \\
SleepSMC$^{c}$ \cite{ma2025sleepsmc}
& 9 & -- & 0.77 & 0.75 & 0.70 \\
SleepSMC, EEG$^{c}$ \cite{ma2025sleepsmc}
& 6 & -- & 0.73 & 0.70 & 0.65 \\
Ours: teacher
& 4 & 136.2M & 0.82 & \textbf{0.81}
& \textbf{0.77} \\
Ours: student, \gtad, 25\%
& 2 & 31k & 0.79 & 0.77 & 0.73 \\
\hline
\multicolumn{6}{l}{\parbox{0.95\columnwidth}{
\scriptsize
$^{a}$C4; patient-wise split not stated.
$^{b}$50 subjects, 25-fold.
$^{c}$5-fold; trained with EEG, EOG and EMG.
}}
\end{tabular}
\end{table}

\textbf{Label efficiency} (Table~\ref{tab:main}).
Trained on all labelled patients, the 31k-parameter students
stage patient sleep well, but their performance drops sharply
across cohorts and input settings when only 25\% of patients
are labelled. The teacher is less affected by label scarcity,
providing informative pseudo-labels for the unlabelled patients.
\gtad\ emphasizes less probable
next-stage assignments while down-weighting transitions
with high between-patient variability. This weighting is designed to help the student learn fragmented sleep patterns without
relying excessively on inconsistent cohort-level transitions.

With the same labels, \gtad\ improves macro F1 over CE by \gain{isruc}{eeg}{gtad_cv}{ce} to
\gain{naps}{eog}{gtad_cv}{ce} points, leaving the students between \gain{isruc}{eog}{gtad_cv}{ce_full} and
\gain{isleeps}{eeg_eog}{gtad_cv}{ce_full} points from their full-label counterparts. All gains are means of
paired per-patient macro-F1 differences, not differences of the fold means in Table~\ref{tab:main}. The largest gains occur in N1,
a transitional stage that is more prevalent in fragmented
clinical sleep and frequently missed by label-scarce CE
students. With EEG+EOG, \gtad\ increases N1 F1 from
\res{isleeps}{eeg_eog}{ce}{f1_N1} to
\res{isleeps}{eeg_eog}{gtad_cv}{f1_N1},
\res{naps}{eeg_eog}{ce}{f1_N1} to
\res{naps}{eeg_eog}{gtad_cv}{f1_N1}, and
\res{isruc}{eeg_eog}{ce}{f1_N1} to
\res{isruc}{eeg_eog}{gtad_cv}{f1_N1}
(iSLEEPS, NAPS, ISRUC-S1), respectively. These results
match or exceed those of the fully supervised CE students
(\res{isleeps}{eeg_eog}{ce_full}{f1_N1},
\res{naps}{eeg_eog}{ce_full}{f1_N1},
\res{isruc}{eeg_eog}{ce_full}{f1_N1}).
Combining EEG and EOG provides the strongest student
performance across the three cohorts.

\textbf{Comparison with published work} (Table~\ref{tab:isruc}): Our teacher has the highest macro F1 on both
cohorts. With 25\% of the labels and 31k parameters, the two-channel \gtad\ student matches the best published macro
F1 on iSLEEPS and, on ISRUC-S1, comes within a few points of models with more channels \cite{zhang2024mssc,ji2022jkstgcn}
while exceeding nine-channel SleepSMC \cite{ma2025sleepsmc}.

\begin{figure}[t]
  \centering
  \includegraphics[width=\linewidth]{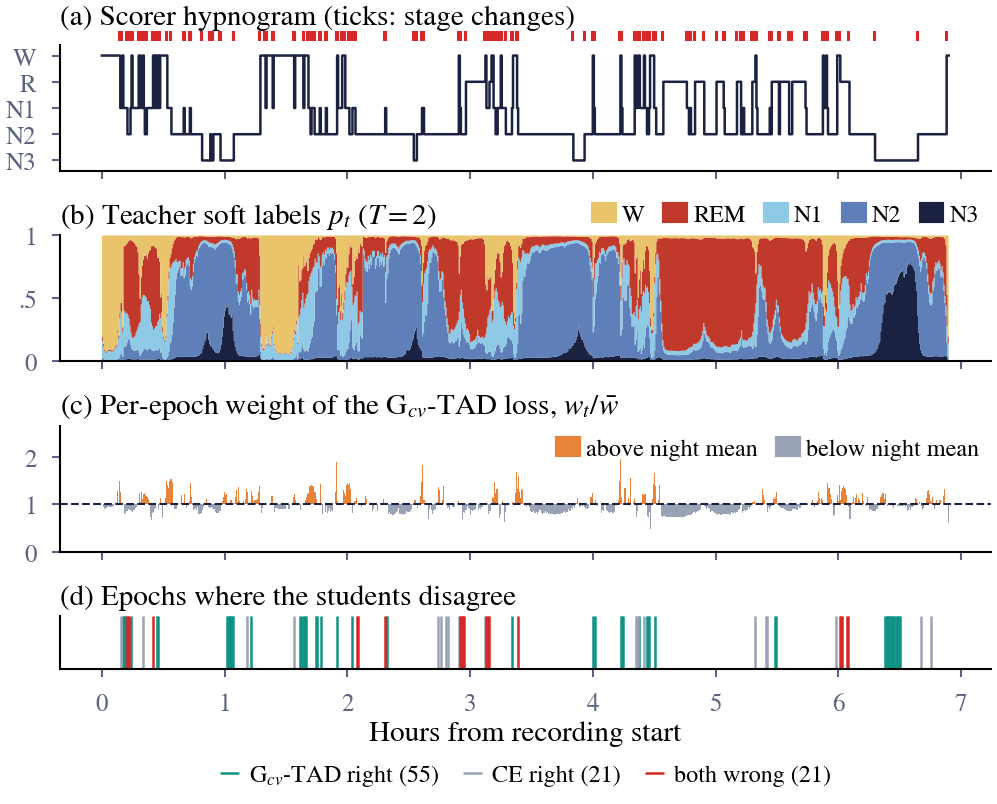}
  \caption{Illustration of \gtad\ on a highly fragmented, unlabelled iSLEEPS training night (fold 1; 16.2 stage changes per hour) (a)~Scorer hypnogram, shown for reference only; red ticks: stage changes. (b)~Teacher soft labels $p_t$ ($T=2$), the distillation
  targets. (c)~Weight of the transition loss per epoch, $w_t=c_{t-1}\sum_k G_{cv}[s,k]\,p_{t,k}$, relative to its
  night mean (dashed: equal weighting). (d)~Epochs where the \gtad\ and CE students (25\% labels) disagree:
  \gtad\ matches the scorer on 55, CE on 21, neither on 21.}
  \label{fig:distill}
\end{figure}

\textbf{Where the transition loss acts:}
Fig.~\ref{fig:distill} illustrates how \gtad\ guides
student learning on a fragmented iSLEEPS training night,
containing 16.2 stage changes per hour compared with
the cohort median of 9.6 among unlabelled patients.
The scorer hypnogram (Fig.~\ref{fig:distill}a) shows
frequent stage changes, while the teacher's soft
predictions (Fig.~\ref{fig:distill}b) remain confident
during consolidated N2, N3 and REM sleep but become
less certain around transitions, particularly N1.
Rather than treating all epochs equally, \gtad\
uses the preceding-stage confidence and a
cohort-specific transition prior to redistribute
the distillation signal. The prior reduces emphasis
on common self-transitions while discounting
stage changes with high between-patient variability.
Consequently, the loss places relatively greater
weight near less frequent stage transitions
(Fig.~\ref{fig:distill}c), where conventional
epoch-wise supervision may provide limited guidance.

This pattern extends across all three cohorts.
In fold-1 training data, stage-change epochs account
for 17, 21 and 29\% of epochs but receive 20, 25
and 41\% of the distillation weight in iSLEEPS,
NAPS and ISRUC-S1, respectively. Similarly,
N1 epochs represent 10, 10 and 13\% of the data
but receive 12, 13 and 21\% of the weight.
The stronger redistribution in ISRUC-S1 is consistent
with its greater sleep fragmentation. This emphasis
on less common transitions and N1 provides a
plausible explanation for the improved N1 staging
observed in the cross-validated results.

The student predictions offer further insight
(Fig.~\ref{fig:distill}d). On the example night,
\gtad\ and the CE student disagree on 97 epochs,
of which \gtad\ correctly classifies 55, compared
with 21 for CE. Across fold-1 training nights,
\gtad\ is correct in 56, 48 and 60\% of
disagreements, versus 27, 29 and 29\% for CE.
However, improvements are not restricted to
transitions; the example also shows gains during
stable N3 sleep. Moreover, the teacher achieves
only 64\% accuracy on this night, illustrating
that distillation remains sensitive to teacher
reliability. These analyses use training recordings
whose scorer labels were hidden during learning
and examined retrospectively. These analyses illustrate how the proposed loss distributes its guidance, while generalization
is evaluated on held-out patients in
Table~\ref{tab:main}.

\textbf{Wearable feasibility:} The student has \eff{student_1ch}{params} parameters, \eff{student_1ch}{ratio}$\times$
fewer than the teacher (136.2M), and needs \eff{student_1ch}{mmacs}M multiply--accumulates per 30-s epoch instead of
\eff{teacher}{gmacs}G. Its weights take \eff{student_1ch}{fp32_kb}~KB in FP32 and \eff{student_1ch}{int8_kb}~KB after
dynamic INT8 quantization, within the SRAM of wearable microcontrollers, and its state between epochs is a single
64-dimensional vector. One epoch takes \eff{student_1ch}{ms}~ms on one thread of an Intel Xeon W-2265 CPU (batch 1)
and is scored as soon as it ends, whereas the teacher needs \eff{teacher}{ms}~ms per epoch even with cached
embeddings and must wait 5~min for future context.

\section{Conclusion}
\label{sec:conclusion}

Long-term sleep monitoring in clinical populations is
limited not only by wearable hardware but also by the
scarcity of expert-labelled patient recordings. Our
findings show that knowledge from a cohort-adapted
foundation model can be transferred to a lightweight,
causal student with substantially fewer labels.
By accounting for differences in sleep transitions
across patients, \gtad\ offers a promising approach
to learning from heterogeneous clinical sleep.
This could enable more accessible longitudinal
monitoring of sleep disruption during recovery
and disease progression, where repeated laboratory
PSG is impractical. Future work will evaluate
wearable electrode configurations, on-device
performance, and generalization to additional
clinical populations before real-world deployment.

\section{Compliance with ethical standards}
\label{sec:ethics}

This retrospective study used de-identified PSG data from three existing clinical datasets: iSLEEPS, NAPS, and ISRUC-S1. The original iSLEEPS study received approval from the NIMHANS Institutional Ethics Committee. NAPS data were accessed through the NSRR under its data use agreement, while ISRUC-S1 is publicly available. 

\section{Acknowledgments}
\label{sec:acknowledgments}
We acknowledge IHub-Data, IIIT Hyderabad (H1-002), for financial assistance.

\section{Declaration of AI Use}
\label{sec:ai_use}

ChatGPT (OpenAI) and Claude (Anthropic) were used solely to improve the quality and clarity of the writing and the aesthetic presentation of the model architecture image. All AI-assisted outputs were reviewed by the authors, who take full responsibility for the content of the paper.
\bibliographystyle{bib/IEEEbib}
\bibliography{bib/refs}

\end{document}